# Collective deceleration: toward a compact beam dump

H.-C. Wu[1], T. Tajima[1,2], D. Habs[1,2], A.W. Chao[3], J. Meyer-ter-Vehn[1]

[1]*Max-Planck-Institut für Quantenoptik, D-85748 Garching, Germany*

[2]*Fakultät für Physik, Ludwig-Maximilians-Universität München, D-85748 Garching, Germany*

[3]*SLAC National Accelerator Center, Stanford University, Stanford, California 94309, USA*

With the increasing development of laser accelerators, the electron energy is already beyond GeV and even higher in near future. Conventional beam dump based on ionization or radiation loss mechanism is cumbersome and costly, also has radiological hazards. We revisit the stopping power of high-energy charged particles in matter and discuss the associated problem of beam dump from the point of view of collective deceleration. The collective stopping length in an ionized gas can be several orders of magnitude shorter than the Bethe-Bloch and multiple electromagnetic cascades' stopping length in solid. At the mean time, the tenuous density of the gas makes the radioactivation negligible. Such a compact and non-radioactivating beam dump works well for short and dense bunches, which is typically generated from laser wakefield accelerator.

## I. INTRODUCTION

As accelerators of particles (electrons and ions) acquire more energies and fluence, the issue of the radiological safety for the operation of such accelerators is increasingly important. At the terminal of every particle accelerator, one needs to decelerate particles into safe energy region, so that there is little radioactivation induced by the high-energy particles in the environment, and no radiation hazard for the laboratory staff. One has to be cautious on the design of the beam dump and radiation shielding. The beam dump perhaps contains both high-Z and low-Z materials and has a thick concrete surrounding. The whole safety system must be validated and monitored. Such a safety issue becomes cumbrous and expensive with the increasing particle energy, especially for the table-top laser wakefield accelerator.

In this paper, we suggest the utilization of the physical consequence of the collective force in matter (in particular in a plasma) when the deceleration becomes dominantly collective. We shall find that short and dense bunches of electrons and other particles such as positrons are amenable under appropriate conditions to their stoppage over many orders of magnitude shorter distance than the conventional beam dump with solid matter. The needed plasma density is low, so hazardous radioactivation due to individual nuclear collisions can be much less than the conventional beam dump. This compact and safe beam dump becomes more pronounced for beyond GeV particles, since some secondary particles can be generated under this energy region, like muons, which are heavy and need a longer distance for stopping in the condensed matter.

In order to make the accelerator and its associated beam dump system compact and safe, we can marshal collective interaction that can far surpass its magnitude over the conventional individual forces if proper conditions are met. In the present article we focus on the deceleration. However, we regard that there can be a general consideration of overall utilization of collective force for the purpose of beam dynamics in order to make the system far more compact over the conventional methods. Using electric (and sometimes magnetic) fields of collective origin in the plasma, one can focus the electron [1] and ion [2] beams. This is so-called plasma lens. Together with the beam dump using the plasma we shall consider below, we can call these efforts as collective plasma optics.

The paper is organized as follows. To compare with the conventional beam dump, in Sec. II, we present a review on the stopping power in matter based on the Bohr-Fermi-Bethe-Bloch theory. In Sec. III, we give the stopping power of the collective deceleration for the dense and ultrashort electron bunch, and compare it with the classic stopping power. PIC simulations on the collective deceleration of electron bunch in the underdense plasma are given in Sec. IV. We find that the deceleration becomes ineffective after a certain distance in the uniform plasma. When that happens, a periodic-structured plasma is proposed to further decelerate the electrons. Moreover, microbunching structure with the period much smaller than the plasma wavelength is developed during the deceleration process. An analytic description is given on the microbunching process and its potential applications are discussed. The final Sec. V draws a conclusion.

## II. BOHR-FERMI-BETHE-BLOCH THEORY

The conventional beam dump is designed based on the understanding of the Bohr-Fermi-Bethe-Bloch classic theory on the stopping power in matter [3-9]. A classic formula given first by Bohr [3] in 1913, and later modified by Bethe [5] and Bloch [6] into a quantum-mechanical formula, which is now universally called the Bethe-Bloch formula [7,8] of the stopping power (for relativistic electrons in condensed matter), reads

$$-(dE/dx)_I = (F/\beta^2)[\ln(2m_e\gamma^2v^2/I) - \beta^2], \qquad (1)$$

where $E$ is the electron kinetic energy, $F = 4\pi e^4 n_{e,m}/m_e c^2 = e^2 k_{pe,m}^2$, $n_{e,m}$ is the electron density in the stopping material, $k_{pe,m} = \omega_{pe,m}/c$ is plasma wavenumber, and $\beta = v/c$ is the normalized electron velocity. The electron energy is converted into the excitation or ionization potential of the bounded electrons in the atom of the stopping material and $I$ represents a specific average of the excitation and ionization potentials in the atom. The dominant mechanism of Bethe-Bloch stopping power is the charged particle interaction with electrons in matter resulting in ionization. The logarithm term within the bracket is around 20 for a broad range of parameters.

In Ichimaru's treatment of the Bethe-Bloch formula in plasma [9], the stopping power is clearly attributed to the part due to the binary collisions and that to the long-ranged collective interaction, where the beam particle is treated as a single test particle. In another word, $-(dE/dx)_P = -(dE/dx)_{ind} - (dE/dx)_{coll}$, where the first term

$$-(dE/dx)_{ind} = (F/\beta^2)\ln(m_e v^2/e^2 k_D) \qquad (2)$$

arises from individual-particle collisions with the characteristic wavenumber limited by $m_e v^2/e^2$ and down to the Debye wavenumber $k_D$; while the second term

$$-(dE/dx)_{coll} = (F/\beta^2)\ln(k_D v/\omega_{pe}) \qquad (3)$$

is the contribution from collectively excited plasma waves with wavenumber $k < k_D$. Interestingly, the plasma stopping power contributions by the individual binary collisions and by the plasma collective oscillations in this linear theoretic regime are of the form that they can be combined together to yield $-(dE/dx)_P = (F/\beta^2)\ln(m_e v^3 / e^2 \omega_{pe})$.

For relativistic electrons, the other important energy loss from individual collisions is due to bremsstrahlung radiation [4] of electrons. In place of Eq.(1) or Eq.(2), the stopping power due to the radiation loss reads

$$-(dE/dx)_R = F(Z/137\pi)(\gamma - 1)\ln(183 Z^{-1/3}), \qquad (4)$$

where $Z$ is atomic number. The approximate ratio of the two losses is $(dE/dx)_R / (dE/dx)_{ind} = EZ/1600 m_e c^2$ [5]. Thus, radiation loss is dominant for higher-energy electrons, e.g. $E > 100 \text{MeV}$ for $Z = 10$. However, emitted radiation quickly spawns multiple generations of cascades of electrons. Because of this, the stopping distance remains considerable.

**III. COLLECTIVE DECELERATION**

The usage of collective fields of plasma for particle acceleration was first suggested by Veksler [10] (deceleration in the present context). It may allow interaction to be enhanced above and beyond the single particle level (Eq.(2)) and the linear level of collective field (Eq.(3)) both for the stooping power and for acceleration. With the development of powerful lasers and high-current relativistic electron bunches, the new method of laser (or plasma) wakefield acceleration has been proposed to accelerate electrons by exploiting collective plasma fields, such as by laser [11,12] and by electron beam [13].

The wakefield amplitude, when driven at resonance of the plasma wave (medium's collective oscillation's eigenfrequency) by strong ponderomotive force of these drivers, becomes highly nonlinear and grows beyond perturbative theory applicability. It may be only characterized by the nonperturbative limit of the wavebreaking field [11] $m_e c \omega_{pe} / e$ and that driven by the electron bunch [14] $m_e c \omega_{pe} (n_b / n_e) / e$ as the wakefield in this limit has a cusp singularity [15], where $n_b$ and $n_e$ are the electron bunch and plasma densities, respectively. This collective stopping power for wakefield deceleration of the electron bunch is large:

$$-(dE/dx)_{coll-wavebreak} = m_e c \omega_{pe} (n_b/n_e). \qquad (5)$$

Note that the linear theoretic wakefield excited by a single test charge in the long wavelength (of the plasma collective field) is given by Eq.(3), as compared with the nonlinear wakefield stopping power Eq.(5).

To optimally generate this non-perturbative plasma wakefield, we require the dense electron bunch $n_b \geq n_e$. To avoid self-injection of plasma electrons, however, we need $n_b \leq n_e$. As a satisfactory compromise, we choose to stay around $n_b/n_e \sim 1$. Thus the ratio of the collective deceleration in plasma and the Bethe-Bloch stopping power in condensed matter is

$$R = \frac{(dE/dx)_{coll-wavebreak}}{(dE/dx)_{ind}} \approx \frac{m_e c \omega_{pe} \beta^2}{F\Lambda} = \frac{n_e}{n_{e,m}} \frac{\lambda_{pe}}{r_0} \frac{\beta^2}{2\pi\Lambda}, \qquad (6)$$

where $\lambda_{pe}$ is plasma wavelength of the background plasma with the density $n_e$, $r_0$ is classical electron radius, and $\Lambda$ is the logarithm term. On the other hand, according to Eq. (2) the ratio of the stopping power due to the individual interaction (short-range) *(dE/dx)$_{ind}$* (in a plasma) to that in the conventional solid dump, *(dE/dx)$_{ind}$* (in a solid) is *n$_e$/n$_{e,m}$*, which is several order of magnitude less than unity. This contributes to the significant reduction of the amount of nuclear activation due to individual nuclear collisions in the plasma dump.

For a typical example with $n_e = 10^{19} \text{cm}^{-3}$, $n_{e,m} = 3 \times 10^{23} \text{cm}^{-3}$, and $\lambda_{pe} = 10 \mu\text{m}$, we have $R \approx 1000$, i.e. the deceleration distance in the underdense plasma is three orders of magnitude smaller than the stopping in condensed matter. Because of this (Eq.(6)) the collective deceleration by a tenuous gas (which will be quickly turned into ionized plasma by the bunch's impinging electric field) is capable of stopping beams by many orders of magnitude shorter (1/R) than the conventional solid beam dump, and yet the radioactivating hazard is reduced by many orders of magnitude (*n$_e$/n$_{e,m}$*) due to the tenuity of the gas compared to the solid and its consequent binary collision scarcity. We simultaneously accomplish the enhancement of the stopping power and the reduction of the binary collisions both by many orders of magnitude.

In relativistic regimes beyond GeV, in addition to the multiple cascades of electrons,

bremsstrahlung photons by the radiation loss generate muon pairs by the photonuclear reaction. The muon fluence is highly peaked in the forward direction. Additional material is needed for stopping them [16]. Due to the heavier mass $m_\mu = 206.8 m_e$, the muon is more penetrative than the electron. The stopping mechanism for muons is the ionization loss. Usually, several meters of high-Z metals are needed to stop the muons. Its stopping power is $-(dE/dx)_{ind,\mu} = -(dE/dx)_{ind}(m_e/m_\mu)$, where $-(dE/dx)_{ind}$ is the electron stopping power given by Eq. (1). Compared with the collective deceleration in the plasma, one has

$$R_\mu = \frac{(dE/dx)_{coll-wavebreak}}{(dE/dx)_{ind,\mu}} = R\frac{m_\mu}{m_e}, \tag{7}$$

where R is given by Eq. (6). In the example of the last paragraph, this ratio takes the value as large as $R_\mu \approx 2 \times 10^5$. Thus, beyond GeV energies the stopping power of the collective deceleration in the plasma is even more pronounced in comparison with the condensed matter Bethe-Bloch taking muons into account.

It is also noted that, recently, collective energy loss of an attosecond electron pulse in overdense plasmas is already discussed in Ref. [17]. Here, we focus on a practical design of a beam dump for electron accelerators.

## IV. PARTICLE SIMULATION

### A. Collective deceleration and saturation

We examine the feasibility of beam dump using collective deceleration in a tenuous plasma, For this we carry out a series of two-dimensional (2D) particle-in-cell (PIC) simulations [18]. As we shall see, our beam dump is highly effective for short and dense beams. Beams from laser wakefield accelerator (LWFA) are in fact very short and dense. Therefore, we take typical parameters of a bunch from LWFA [19]. The electron bunch has a total charge 50-100pC, and a spherical distribution with diameter of $3\mu m$. The bunch density is $n_b \approx 2.2 - 5 \times 10^{19} \text{cm}^{-3}$. The beam divergence angle is $\theta = 1\text{mrad}$, which is used to calculate the bunch size after a certain vacuum drift. As shown in Ref. [20], such an ultrashort bunch mainly experiences transverse expansion. The energy spread is 1%. The electron bunch is injected into the plasma and propagates from left to right along the x axis. The simulation box size is $10\lambda_{pe} \times 10\lambda_{pe}$, and it moves with the light speed.

Figure 1 shows the bunch total energies as a function of the propagation distance in the uniform plasma with the density $n_e = n_b/5 \approx 1.1 \times 10^{19} \text{cm}^{-3}$. In this case, the normalized transverse size and longitudinal length are $\sigma_T/\lambda_{pe} = \sigma_L/\lambda_{pe} = 0.3$. The initial particle energy $E$ is varied from 100 MeV to 100 GeV and $E_{\text{GeV}}$ is the bunch energy in GeV unit. Over a broad range of energies, the energy losses as a function of energy are isomorphic. For the 1 GeV case, 75% energy is deposited in the 1.5 mm long plasma. The deceleration distance is proportional to the bunch energy, and in fact determined by the stopping power, as shown in Eq.(5), i.e. wakefield amplitude is independent of the bunch energy, until saturation. We define the saturation length $L_s$. After the saturation length, the electron deceleration becomes much slower and almost vanishes.

To understand saturation mechanism, Fig. 2 provides the behavior of energy vs. x position of all electrons in the bunch around the distance $L_s$. Here, we take the initial bunch energy of 500 MeV and the plasma density $n_e = 2n_b \approx 4.4 \times 10^{19} \text{cm}^{-3}$. In this case the normalized bunch sizes are $\sigma_T/\lambda_{pe} = \sigma_L/\lambda_{pe} = 0.6$. Figure 2(a) shows the bunch tail is effectively decelerated and in Fig. 2(b) some tail electrons are completely stopped toward zero velocity and lag behind the main bunch. Then, these lagging electrons are trapped in the acceleration phase of the wakefield and regain their energy, as shown in Fig. 2(c). In fact, the bunch is already split into three parts and the deceleration saturates in Fig. 2(c). The total energy evolution is shown in Fig. 3(b). The remaining energy after saturation is about 25%.

**B. Beam dump with structured plasmas**

To circumvent the saturation in uniform plasma and further decelerate the bunch, we suggest to employ a structured plasma for phase mismatch control [21] as shown in Fig. 3(a) and Fig .4(a). Just before the moment when some tail electrons are completely stopped, we replace the uniform plasma by some periodic plasma slabs with vacuum gaps or inserted periodic thin foils in the background uniform plasma. It is expected that those electrons which approach to come to rest can be retained around the vacuum gap or foil, and not trapped in the plasma for acceleration.

In the vacuum gap case, we set the length of the plasma slab the same as the vacuum gap. Figure 3(b) shows the bunch energy can indeed continue to decelerate further through introduction of the structured plasma with the periods $L_P/\lambda_{pe}=2$, 5 and 10 after a deceleration in the 1.15 mm long uniform plasma. After a distance of 3mm, 90% bunch energy is absorbed. Further deceleration is possible if longer structured plasma is used.

As expected, Fig. 3(c) shows that a low-energy electron tail is left after the main bunch. Most of these low-energy electrons have a kinetic energy smaller than 5 MeV. For the electrons less than 10 MeV, they are safer and may not lead to radioactivation. The bunch head cannot effectively be decelerated, because the wakefield is weak on the bunch head.

To check the robustness of the deceleration in the structured plasma, we consider an electron bunch after a 1 cm vacuum drift. The bunch transverse size becomes $\sigma_T=10\mu\text{m}$, and longitudinal length does not change with $\sigma_L=3\mu\text{m}$. The bunch density is $n_b\approx 2\times 10^{18}\text{cm}^{-3}$. We vary the plasma density from $n_e/n_b=1$ to $n_e/n_b=80$. The corresponding normalized bunch length is variational from $\sigma_L/\lambda_{pe}=0.1$ to $\sigma_L/\lambda_{pe}=1.2$ and bunch width from $\sigma_T/\lambda_{pe}=0.4$ to $\sigma_T/\lambda_{pe}=3.8$. For the uniform plasma case, the loss rate of the bunch energy decreases with the normalized bunch length as shown in Fig. 3(d). This is because that for a relatively longer bunch (i.e. shorter plasma wavelength) with $\sigma_L/\lambda_{pe}\sim 1$, the bunch tail is always accelerated and the wakefield excitation is not optimal [14]. The optimal wakefield is generated for $\sigma_L/\lambda_{pe}=0.5$. In another word, the proposed beam dump is effective when

$$\sigma_L/\lambda_{pe}<1. \quad (8)$$

This implies that (i) the shorter the bunch is, the higher the plasma density may be taken and the shorter the stopping length becomes (see Eq. (5)) and that (ii) the denser bunch is more effective (also Eq. (5)). For the structured plasma case the energy loss rate can improve to 90% and be independent of the normalized bunch length. The results in Fig. 3 show that the structured plasma with the vacuum gap can be applicable and robust for a broad range of bunch or plasma parameters.

Figure 4(a) shows the other structured plasma scheme with periodic thin foils dipping in the uniform plasma. This scheme may be easier in the experiment aspect. With this method, like Fig. (3), we improve the deceleration efficiency after a deceleration in the 1.15 mm long uniform plasma. The length and separation of foils are $0.1\lambda_{pe}$ and $L_P = 1\lambda_{pe}$, respectively. The foil density is 100 times of the background plasma, i.e. $n_{foil} \approx 4.4 \times 10^{21} \text{cm}^{-3}$, which is typical density of the solid aerogel. As shown in Fig. 4(b), after a distance of 3mm, 85% bunch energy is absorbed. Figure 4(c) shows a low-energy electron tail after the main bunch. Most of these low-energy electrons have a kinetic energy smaller than 10 MeV. Further deceleration is possible for longer structured plasma or optimized parameters of foil densities and thickness.

In addition, we have also examined the case of positron beam deceleration. We have found that positron bunches may be as well decelerated as electron bunches from our simulation.

## C. Microbunching of the decelerated bunch

Simulations also show the electron bunch can develop refined microbunch structure during the collective deceleration. This is illustrated in Fig. 5 for the case of Fig. 2. The electron bunch carries out betatron oscillations in the transverse direction. The modulation period of the microbunch structure decreases with the propagation distance. The reason for the microbunch generation is due to the nonuniform radial wakefield along the longitudinal direction within the bunch. The electron bunch can be considered as a set of infinitely thin sheets along the x direction. If the radial field is uniform along the x direction, the radius of each sheets oscillate synchronously with the same betatron frequency. For our case, the wakefield is weak towards the bunch head and is strong towards its tail. The different sheets therefore have different betatron frequencies and the resulting nonsynchronous oscillations lead to the bunch envelope modulation.

Since the beam deceleration works near the blow-out regime, we assume that electron bunch blows out all the plasma electrons, and leaves an positive ion column. The transverse electrostatic field of the ion column is $2\pi n_e e r$. The electron carries the transverse betatron motion $dp_T/dt = -2\pi n_e e^2 r$ in this transverse field. For relativistic electrons, we have $\gamma_T = 1/\sqrt{1-v_T^2} \ll \gamma$. The electron motion equation becomes

$$d^2 r/dt^2 = -\Omega_b^2 r, \tag{9}$$

where $\Omega_b = \omega_{pe}/\sqrt{2\gamma}$ is betatron frequency. Since a relativistic electron has $t \simeq x/c$, the motion equation is rewritten as $d^2 r/dx^2 + (\Omega_b^2/c^2)r = 0$, where x is the electron propagation distance.

If we neglect the effects of emittance, space charge, and self-magnetic field of the electron bunch, from Eq. (9), we can obtain the envelope equation of the bunch [22,23] as

$$[\frac{\partial^2}{\partial x^2} + \frac{\Omega_b^2(\xi)}{c^2}]\sigma_T(x,\xi) = 0, \tag{10}$$

where $\xi = x - ct$ is the co-moving coordinate of the bunch. We consider the front part of electron bunch within $\xi \in [-\sigma_L', 0]$, where $\sigma_L' \leq \sigma_L$. We assume the radial field increases linearly from the bunch head $\xi = 0$ to the position $\xi = -\sigma_L'$, so one has

$$\Omega_b(\xi) = \Omega_{b0}(1 + \xi/\sigma_L'), \tag{11}$$

where $\Omega_{b0} = \omega_{pe}/\sqrt{2\gamma}$ is the maximum betatron frequency. The solution of Eq. (10) is

$$\sigma_T(x,\xi) = \sigma_T(0,\xi)|\cos[\Omega_{b0}(1+\xi/\sigma_L')x/c]|. \tag{12}$$

The modulation period of the bunch envelope as a function of $\xi$ is

$$\eta = \frac{\pi c \sigma_L'}{\Omega_{b0} x} = \sqrt{\frac{\gamma}{2}} \frac{\sigma_L'}{x} \lambda_{pe}, \tag{13}$$

which decreases with the propagation distance x. For the case of Fig. 5 we find $\sigma_L' \approx 0.5 \lambda_{pe}$. Substituting $\gamma = 1000$ and $\lambda_{pe} = 5\mu m$, we obtain $\eta_{0.5mm} \approx 0.56 \mu m$ and $\eta_{0.8mm} \approx 0.35 \mu m$, which agree with the median in Fig. 5(c). The chirped structure in Fig.

5(c) is due to the nonlinear wakefield rising within the bunch.

Such a microbunched electron bunch can potentially be a source for coherent radiation or to feed a free electron laser, and its generation requires only a short plasma insertion. Of course, additional investigations on optimum microbunch generation are needed for a practical application in this direction. We notice that there remains some chirp in the period of the microbunches. Since we understand the reason for this in the nonlinear chirp of the betatron frequency, we can utilize this or control it. It is also possible to using this new microbunching mechanism to generate zeptosecond electron pulse train from an attosecond bunch as the status in Ref. [17]. Zeptosecond electron pulse train can be used for a diagnosis tool for ultrafast phenomenon in atom and nuclear physics.

## V. CONCLUSION

In conclusion, we have suggested the utilization of the collective deceleration in plasma as a beam dump mechanism for electron accelerators. This new method provides some 3-5 orders of magnitude more efficient beam dump capability over the conventional beam dump, and reduces radioactivated hazard by many orders of magnitude. It could dramatically decrease the cost of the beam dump and relax the severity of the radiation shielding. The conditions necessary for effective collective deceleration call for short and dense beams. These conditions are ideally match with the beam characteristics of laser electron accelerators., Thus this technique could eventually benefit the development of the high-energy laser particle accelerator system in quite a unique fashion. Together with other collective plasma optics, we may design compact accelerator and its associated systems with the present collective decelerator.

In principle, the deposited energies from the decelerated beams in the form of organized plasma wakefields, unlike the heat energy in a conventional dump, may be recovered into electricity [24]. This may be return to drive the accelerator, saving the energy. In addition, the microbunching mechanism clarified in this paper may provide a new method for the seeding of the free electron laser and ultrashort (zeptosecond) bunch train generation.

**ACKNOWLEDGMENTS**

This work was supported by Deutsche Forschungsgemeinschaft through the DFG-Cluster of Excellence Munich-Centre for Advanced Photonics (MAP) and Transregio TR18. Wu acknowledges support from the Alexander von Humboldt Foundation.

Figure 1

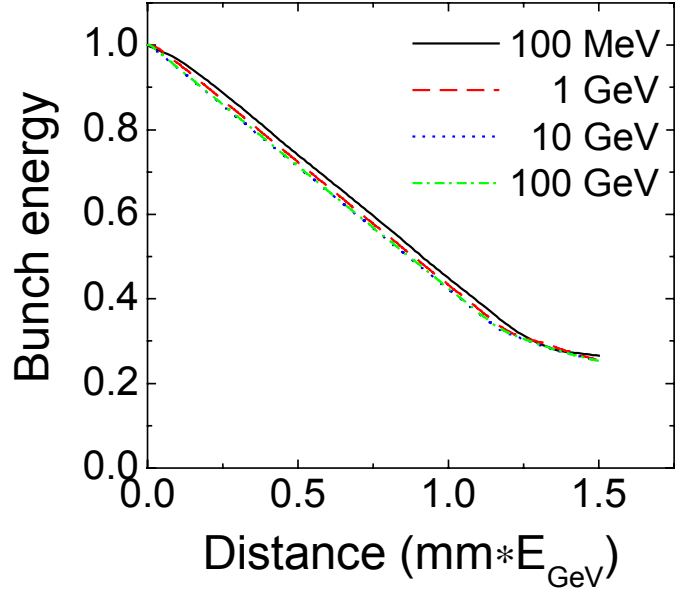

FIG. 1. (Color) Collective stopping power rate and its saturation. The evolution of the bunch total energy normalized to the initial energy with the propagating distance. We take the plasma density $n_e = n_b/5 \approx 1.1\times10^{19}\,\text{cm}^{-3}$ and four initial bunch energies: 100 MeV, 1GeV, 10GeV and 100GeV. $E_{\text{GeV}}$ is the bunch initial energy in GeV unit.

# Figure 2

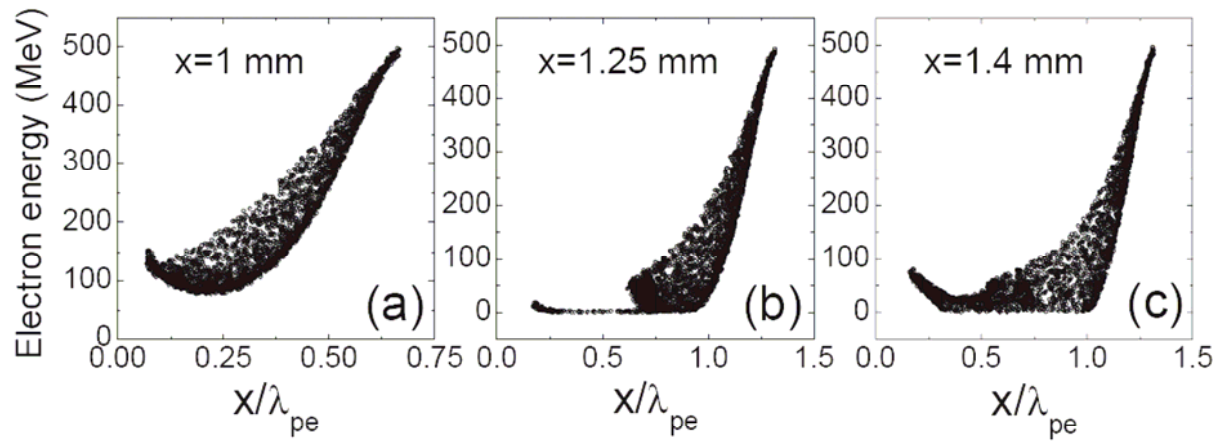

FIG. 2. The collective deceleration and saturation mechanism. The electron energy vs. longitudinal (x) position at the different propagating distances x=1 mm (a), x=1.25mm (b), and x=1.4 mm (c). We take the plasma density $n_e = 2n_b \approx 4.4 \times 10^{19} \text{cm}^{-3}$ and the initial bunch energy 500 MeV.

# Figure 3

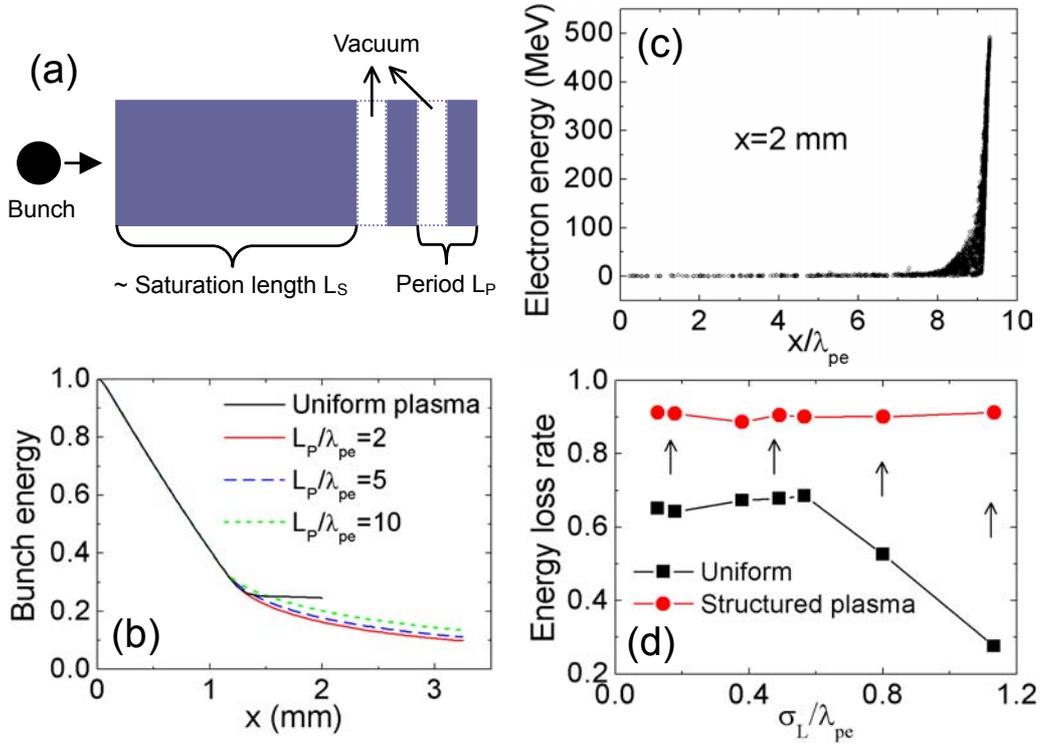

FIG. 3. (Color) Suggested beam dump. (a) The proposed structured plasma consisting of periodic plasma slabs with the vacuum gap. $L_p$ is period of the structured plasma. In each period, the plasma slab length is equal to the vacuum length. (b) The bunch energy evolution for both uniform plasma and the structured plasma with $L_P/\lambda_{pe} = 2$, 5 and 10. (c) The electron energy vs. x position at x=2 mm for the $L_P/\lambda_{pe} = 2$ case. (d) The improvement of the energy loss rate by the structured plasma for the bunch with parameters: $\sigma_T = 10\,\mu m$ and $\sigma_L = 3\,\mu m$. One adjusts the background plasma density to change $\sigma_L/\lambda_{pe}$.

# Figure 4

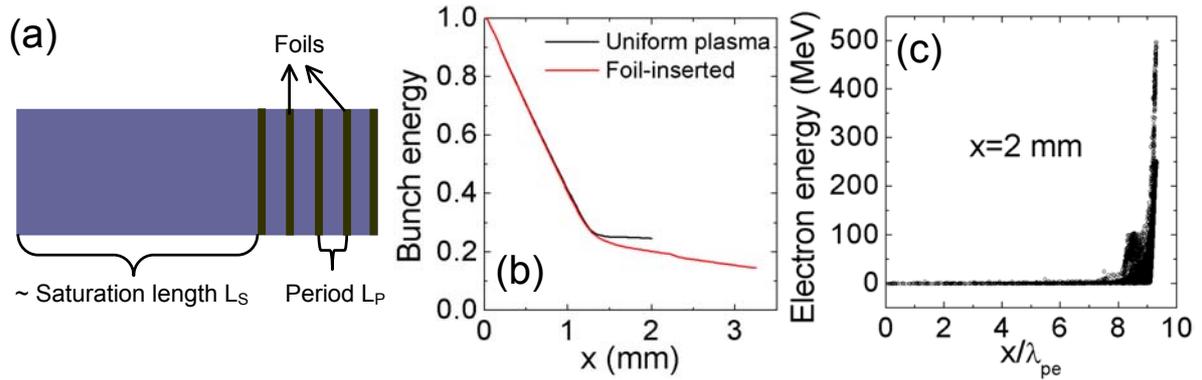

FIG. 4. (Color) Suggested beam dump. (a) The structured plasma with inserted periodic thin foils. $L_p$ is the separation of the neighboring foils. (b) The bunch energy evolution for both uniform plasma and the structured plasma with $L_P/\lambda_{pe}=1$, foil length $0.1\lambda_{pe}$ and density $n_{foil}=100n_e$. (c) The electron energy vs. x position at x=2 mm.

Figure 5

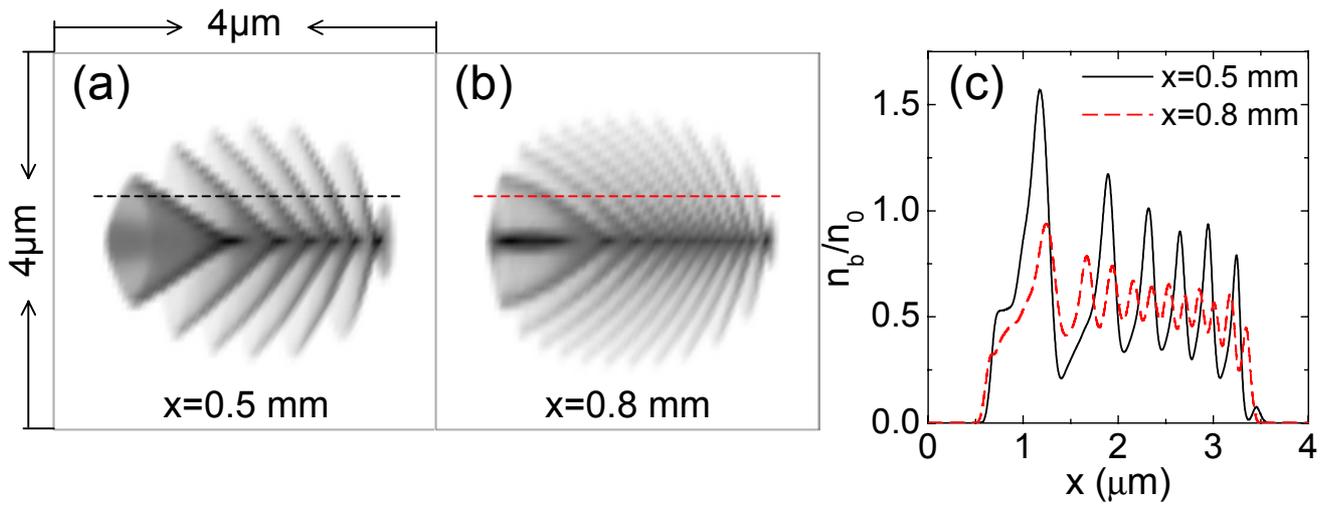

FIG. 5. (Color) Microbunching during deceleration. Snap shots of bunch density for the propagation distances (a) x=0.5 mm and (b) x=0.8 mm. (c) displays the bunch density distributions along the dashed lines in (a) and (b). Simulation parameters are same as Fig. 2.